\documentclass[12pt]{article}
\usepackage{graphicx}

\usepackage{scicite}

\usepackage{times}

\newenvironment{sciabstract}{%
\begin{quote} \bf}
{\end{quote}}

\newcounter{lastnote}
\newenvironment{scilastnote}{%
\setcounter{lastnote}{\value{enumiv}}%
\addtocounter{lastnote}{+1}%
\begin{list}%
{\arabic{lastnote}.}
{\setlength{\leftmargin}{.22in}}
{\setlength{\labelsep}{.5em}}}
{\end{list}}

\title{Quantum Coherence in an Exchange-Coupled Dimer of Single-Molecule Magnets}

\author
{S. Hill,$^{1\ast}$ R. S. Edwards,$^{1}$ N. Aliaga-Alcalde$^{2}$
and G. Christou$^{2}$
\\
\\
\normalsize{$^{1}$Department of Physics, University of Florida,
Gainesville, FL 32611,USA,}\\ \normalsize{$^{2}$Department of
Chemistry, University of Florida, Gainesville, FL 32611, USA.}\\
\\
\normalsize{$^\ast$To whom correspondence should be addressed;
E-mail:  hill@phys.ufl.edu.} }

\date{}

\begin{document}


\baselineskip24pt


\maketitle


\begin{sciabstract}
A multi-high-frequency electron paramagnetic resonance method is
used to probe the magnetic excitations of a dimer of
single-molecule magnets. The measured spectra display well
resolved quantum transitions involving coherent superposition
states of both molecules. The behavior may be understood in terms
of an isotropic superexchange coupling between pairs of
single-molecule magnets, in analogy with several recently proposed
quantum devices based on artificially fabricated quantum dots or
clusters. These findings highlight the potential utility of
supramolecular chemistry in the design of future quantum devices
based on molecular nanomagnets.
\end{sciabstract}

\clearpage
\paragraph*{}

Considerable effort has focused on finding building blocks with
which to construct the quantum logic gates (qubits) necessary for
a quantum computer \cite{QCreview,QCBook}. Most proposals
utilizing electronic spin states take advantage of
nano-fabrication methods to create artificial molecules, or
magnetic quantum dots \cite{Qdot1,Qdot2}. A Heisenberg-type
exchange coupling between dots is achieved by allowing the
electronic wavefunctions to leak from one dot to the next. It is
this coupling which is the essential ingredient in a quantum
device because, unlike classical binary logic, it enables encoding
of data via arbitrary superpositions of pure quantum states, e.g.
$|0\rangle$ and $|1\rangle$ \cite{QCBook}. These superposition
states can store information far more efficiently than a classical
binary memory. Furthermore, they permit massively parallel
computations, i.e. many simultaneous quantum logic operations may
be implemented on a single superposition state. For a quantum
device to become a viable technology, it should be possible to
perform a reasonably large number of quantum operations
($\sim10^4$) on a single qubit without the superposition states
losing phase coherence. Herein lies one of the main technical
challenges, as most quantum systems are highly susceptible to
decoherence through coupling to their environment
\cite{Decoherence}.

We demonstrate that single-molecule magnets (SMMs) may be
assembled to form coupled quantum systems of dimers (or chains,
etc.), with many of the attributes of quantum-dot-based schemes.
Most importantly, our electron paramagnetic resonance (EPR)
investigations of crystals (large, highly ordered 3D arrays)
containing exchange-coupled dimers of SMMs show that decoherence
rates are considerably less than the characteristic quantum
splittings ($\Delta/h\sim$~GHz, where $\Delta$ is the energy
splitting and $h$ is the Planck constant) induced by the exchange
couplings within the dimers, representing a step forward in the
drive towards potential applications involving molecular magnets.
Several proposals have suggested possible quantum computing
schemes utilizing molecular magnets~\cite{Leuen,oxfordQC,Meier}.
The supramolecular (or "bottom-up") approach to materials design
is particularly attractive, as it affords control over many key
parameters required for a viable qubit: simple basis states may be
realized through the choice of molecule; exchange couplings may
then be selectively designed into crystalline arrays of these
molecules; finally, one can isolate the qubits to some degree by
attaching bulky organic groups to their periphery.

The subject of this investigation is the compound
[Mn$_4$O$_3$Cl$_4$(O$_2$CEt)$_3$(py)$_3$]$_2\cdot$2C$_6$H$_{14}$
(hereafter [Mn$_4$]$_2$; EtCO$_2^-$ is propionate, py is pyridine,
and C$_6$H$_{14}$ is hexane) \cite{Mn4}, a member of a growing
family of Mn$_4$ complexes which act as SMMs~\cite{MRS,Angew},
having a well defined ground state spin of $S=\frac{9}{2}$. This
compound crystallizes in a hexagonal space group (R$\bar{3}$) with
the Mn$_4$ molecules lying head-to-head on a crystallographic
S$_6$ axis. The resulting [Mn$_4$]$_2$ supramolecular dimer is
held together by six weak C$-$H$\cdot\cdot\cdot$Cl hydrogen bonds
(Fig.~1), leading to an appreciable antiferromagnetic
superexchange coupling ($J\sim 10~\mu$eV) between the Mn$_4$ units
within the dimer, which influences the low-temperature quantum
properties of related [Mn$_4$]$_2$ dimers~\cite{WernsNature}. Like
all SMMs, [Mn$_4$]$_2$ displays superparamagnetic-like behavior at
high temperatures, and magnetic hysteresis below a characteristic
blocking temperature ($\sim1$~K). The hysteresis loops exhibit
steps, which are due to magnetic quantum tunneling (MQT). However,
unlike isolated SMMs, there is an absence of MQT at zero-field,
due to a static exchange bias field which each molecule
experiences due to its neighbor within the
dimer~\cite{WernsNature}. The effect of the bias is to shift the
field positions of the main MQT steps by an amount of order
$-JS^2/\mu$ (where $\mu$ is the magnetic moment of a Mn$_4$
monomer), so that the first step is observed on the hysteresis
loop before reaching zero-field. However, the exchange bias by
itself does not quantum mechanically couple the SMMs within the
dimer.

Before presenting experimental evidence for the coupled nature of
the dimers, we develop a quantum mechanical model which takes this
coupling into account. Neglecting off-diagonal crystal field terms
and inter-molecular interactions, the effective spin Hamiltonian
(to fourth order) for a magnetic field ($B_z$) applied parallel to
the easy ({\em z}-) axis of a single isolated SMM has the
form~\cite{Angew}

\begin{equation}
\label{Eq1} \hat{H}_i = D\hat{S}^2_{zi} + B_4^0\hat{O}_4^0 +
g_{z}\mu _B B_z \hat{S}_{zi},\end{equation}

\noindent{where $\hat{S}_{zi}$ is the {\em z}-axis spin projection
operator, and the index $i$ ($=1,2$) is used to label the two
Mn$_4$ molecules in the dimer for the interacting case below; {\it
D} ($<0$) is the uniaxial anisotropy constant; $B_4^0\hat{O}_4^0$
characterizes the fourth order axial anisotropy; and $g_z$ is the
{\em z}-component of the Land$\acute{e}$ g-tensor. The omission of
transverse terms in Eq.~\ref{Eq1} does not affect the EPR spectra
(they merely result in weak avoided level crossings which cause
the MQT).}

For the case of two quantum mechanically coupled SMMs, the
effective dimer Hamiltonian ($\hat{H}_D$) may be separated into
the following diagonal and off-diagonal terms

\begin{equation}
\label{Eq2} \hat{H}_D = [\hat{H}_1 + \hat{H}_2 +
J_z\hat{S}_{z1}\hat{S}_{z2}] + \{{\textstyle{1 \over
2}}J_{xy}(\hat{S}_1^+\hat{S}_2^- + \hat{S}_1^-\hat{S}_2^+)\},
\end{equation}

\noindent{where $\hat{H}_1$ and $\hat{H}_2$ are given by
Eq.~\ref{Eq1}, the cross terms describe the exchange coupling
between the two SMMs within the dimer, and the $J$ values
characterize the strength of this coupling. The diagonal zeroth
order Hamiltonian ($\hat{H}_{0D}$, in square brackets) includes
the exchange bias $J_z\hat{S}_{z1}\hat{S}_{z2}$ which has been
considered previously \cite{WernsNature}. The zeroth order
eigenvectors for the dimer may be written as products of the
single-molecule eigenvectors $|m_1\rangle$ and $|m_2\rangle$
(abbreviated $|m_1,m_2\rangle$), where $m_1$ and $m_2$ represent
the spin projections of the two molecules within the dimer. The
zeroth order eigenvalues are then easily obtained by solving Eq.~1
separately for molecules 1 and 2, and adding the exchange bias
$J_z m_1 m_2$ (Fig.~2).

In EPR, the only effect of the exchange bias is to cause shifts in
the positions (energies) of single-spin transitions
($m_i\rightarrow m_i\pm 1$), with the magnitude of the shift
(bias) depending on the state $m_j$ of the other molecule within
the dimer. It is the off-diagonal interaction in Eq.~\ref{Eq2}
($\hat{H}^\prime_D$ in curly brackets) which couples the
molecules, giving rise to the possibility of single-photon
transitions between coupled states of the dimer. In principle, one
could observe this coupling in hysteresis measurements, as
magnetic relaxation mediated by tunneling into distinct
superposition states occurs at slightly different magnetic field
strengths, even when the tunneling occurs via states with the same
total spin projection ($=m_1+m_2$). However, the predicted
splittings of MQT resonances turn out to be less than the
inhomogeneous linewidths of the hysteresis steps
\cite{WernsNature}. Thus, clear evidence for the coupled nature of
the dimer system has so far been lacking.

For illustrative purposes, we treat $\hat{H}^\prime_D$
perturbatively. As this interaction conserves angular momentum,
the eigenvectors may be grouped into multiplets based on the sum
of the projections M~$=m_1+m_2$. $\hat{H}^\prime_D$ then acts only
between states within a given multiplet. The zeroth order
eigenvectors are grouped according to this scheme in Fig~2 (left).
In first order, $\hat{H}^\prime_D$ acts between zeroth order
eigenvectors $|m_1,m_2\rangle$ and $|m_1\pm 1,m_2\mp 1 \rangle$.
The effect of this first order interaction is most apparent in the
M~$=-8$ multiplet, where it lifts the degeneracy between the
$|-\frac{9}{2},-\frac{7}{2}\rangle$ and
$|-\frac{7}{2},-\frac{9}{2}\rangle$ states. The resultant
eigenvectors correspond to symmetric ($S$) and antisymmetric ($A$)
superpositions of the original product states. Indeed,
$\hat{H}^\prime_D$ causes considerable mixing of the zeroth order
eigenvectors within all multiplets, resulting in the first order
corrected eigenvectors which are listed in Fig.~2 (right) for the
lowest four multiplets; here, $|m_1,m_2\rangle_S$ implies
$(|m_1,m_2\rangle+|m_2,m_1\rangle)$ and $|m_1,m_2\rangle_A$
implies $(|m_1,m_2\rangle-|m_2,m_1\rangle)$.


In Fig.~2, we display a schematic of the energy level shifts and
splittings (not to scale) caused by the exchange bias, and by the
full exchange, for the lowest lying levels at high magnetic fields
(M~$=-9$ to $-6$). The states are numbered for convenient
discussion of the data. For clarity, higher lying states with
M~$>-6$, including the zero-field
$|\pm\frac{9}{2},\mp\frac{9}{2}\rangle$ ground states, are not
shown in Fig.~2. Application of a magnetic field parallel to the
easy axis merely shifts all of the zeroth order levels by an
amount $g\mu_B B_z$M. Thus, $\delta$M$=\pm1$ EPR transition matrix
elements may be accurately calculated using the eigenvectors in
Fig.~2. The magnetic dipole perturbation only allows transitions
between states having the same symmetry. The strongest of these
transitions are shown in Fig.~2, labeled (a) through (g).

In the left-hand panel of Fig.~3, we display temperature dependent
high-frequency EPR spectra obtained at 145~GHz, with the magnetic
field applied parallel to the easy ($z$-) axis of a small
($<1$~mm$^3$) single-crystal sample; details concerning our
high-frequency EPR setup are given elsewhere~\cite{MolaRSI}. The
inset shows a single 6~K spectrum ($f=140$~GHz) for a related
monomeric Mn$_4$ complex without head-to-head
interactions~\cite{HillMn4Poly}. The monomer data are typical of
most SMMs, showing a series of more-or-less evenly spaced
resonances, and a smooth variation in intensity from one peak to
the next. By contrast, the dimer spectra exhibit considerable
complexity. In spite of this, the simulated dimer spectra (colored
traces in the right-hand panel of Fig.~3) show remarkable
agreement with the raw data, both in the peak positions and
relative intensities. The optimum parameters were deduced from a
single fit to Eq.~\ref{Eq2} of the main EPR peak positions
obtained at many microwave frequencies. This fit, displayed in
Fig.~4, yields the following values: $D=-0.750(15)$~K,
$B^0_4=-5(2)\times10^{-5}$~K, $g_z=2$ and $J=0.12(1)$~K. These
crystal field parameters are very similar to those obtained for
the monomer [$D=-0.7$~K,
$B^0_4=-9\times10^{-5}$~K~\cite{HillMn4Poly}]. We did not find it
necessary to include anisotropy in the superexchange coupling for
the dimer (i.e. $J_z=J_{xy}=J$), though long range dipolar
interactions improved the quality of the fit~\cite{Park1,dipolar}.

The simulated spectra (Fig.~3) are mainly limited to transitions
among the levels displayed in Fig.~2 [(a) through (g)]; we have
also included the
$(7)_{S,A}\rightarrow|-\frac{9}{2},-\frac{1}{2}\rangle$ and
$|-\frac{9}{2},-\frac{1}{2}\rangle\rightarrow|-\frac{9}{2},+\frac{1}{2}\rangle$
transitions, labeled (h) and (i) respectively. Resonance ($x$),
meanwhile, corresponds to the degenerate
$|+\frac{9}{2},-\frac{9}{2}\rangle\rightarrow|+\frac{9}{2},-\frac{7}{2}\rangle$
and
$|-\frac{9}{2},+\frac{9}{2}\rangle\rightarrow|-\frac{7}{2},+\frac{9}{2}\rangle$
transitions. The only significant differences between the
experimental data and simulated spectra are seen in the $2-3$~T
region, which is due to fact that we did not consider several
moderately strong transitions involving higher lying (M~$>-6$)
states. We deliberately avoid reference to superposition states in
discussing resonance ($x$), as the interaction between the
$|\pm\frac{9}{2},\mp\frac{9}{2}\rangle$ states is extremely weak
(9th~order in $\hat{H}^\prime_D$). Consequently, even the weakest
coupling to the environment would likely destroy any coherence
associated with the
$2^{-1/2}|+\frac{9}{2},-\frac{9}{2}\rangle_{S,A}$ superposition
states.


Resonance ($x$) is observed only over a narrow low-field region
($<0.7$~T) over which the $|\pm\frac{9}{2},\mp\frac{9}{2}\rangle$
levels represent the ground states of the dimer. By following the
relative intensities of resonances ($x$) and (a), one obtains an
independent thermodynamic estimate of the exchange bias which is
in excellent agreement with the value obtained above, and with
independent hysteresis measurements for the same
complex~\cite{privcom}. We note that the previously published
measurements of the exchange bias in [Mn$_4$]$_2$ involved a
slightly different solvent of crystallization, the full compound
having the form
[Mn$_4$O$_3$Cl$_4$(O$_2$CEt)$_3$(py)$_3$]$_2\cdot$8MeCN~\cite{Mn4,WernsNature}.
EPR studies for this complex~\cite{HillMn4Poly} show fewer
transitions from excited levels [transitions ($x$), (a), (b) and
(c) remain clearly visible]. Nevertheless, one can still estimate
a coupling constant [$J=0.10(1)$~K] from the exchange bias, which
is in agreement with the published value~\cite{WernsNature}.


The inset to the right panel of Fig.~3 shows that it is the
transverse part of the exchange ($\hat{H}^\prime_D$) which brings
the simulations into excellent agreement with the data. Indeed,
there is no way to obtain anything closely resembling the
experimental data without including $\hat{H}^\prime_D$ in the
calculation, thus providing compelling evidence that the molecules
are coupled quantum mechanically. The issue of quantum coherence
is best illustrated by examining the splitting of resonances (f)
and (g) $\--$ this splitting is directly proportional to $J_{xy}$,
and corresponds to the $\sim9$~GHz shift of the $(4)_S$ level
relative to $(5)_A$ (Fig.~2). If the phase decoherence rate
($\tau_\phi^{-1}~\equiv$ characteristic rate associated with the
collapse of a quantum mechanical superposition state) were to
exceed 9~GHz, one would expect broad EPR peaks due to transitions
between bands of incoherent states; these bands would occupy the
gaps between the energies given by the exchange bias picture and
the full exchange calculation in Fig.~2, thereby smearing out most
of the sharp features in the observed spectrum. In principle,
$\tau_\phi$ is the same as the transverse spin relaxation time
T$_2$, which can be estimated from EPR linewidths
($\Delta$M~$=\pm1$ transitions)~\cite{T2}. However, we know that
these widths are dominated by weak dimer-to-dimer variations
(strains) in the Hamiltonian parameters, i.e. the actual
$\tau_\phi^{-1}$ is buried within the inhomogeneous EPR
linewidths~\cite{HillMn4Poly,Park1,dipolar}, and is probably much
less than 9~GHz. As a worst case, the narrowest EPR lines would
imply a decoherence time on the order of 1~ns. In order to
determine the real T$_2$ ($\equiv\tau_\phi$), one should carry out
time resolved (pulsed) EPR experiments, e.g. the
free-induction-decay of an initially saturated EPR transition, or
Rabi spectroscopy~\cite{T2}. Time resolved experiments in this
frequency range are technically challenging but, nevertheless,
represent a future objective.


The magnitudes of the quantum splittings (in frequency units)
provide a rough estimate of the rates at which one could perform
computations. In comparison to many competing technologies [e.g.
NMR~\cite{NMR}] these rates are high for electronic spin states,
i.e. GHz rather than kHz or MHz. The largest quantum splittings
($\Delta/h$) for the dimer are on the order of a few tens of GHz.
In fact, $\Delta\tau_\phi/h$ represents a rough figure of merit
for a quantum device, as it gives an estimate of the number of
qubit operations one could perform without loss of phase
coherence. For the worst case given above, $\Delta\tau_\phi/h\sim
30-100$; in reality, it may well be $10^4$ or greater. The most
useful coupled states of the dimer would be the antiferromagnet
zero-field $2^{-1/2}|+\frac{9}{2},-\frac{9}{2}\rangle_{S,A}$
ground states, or Bell states~\cite{QCBook}. As already discussed,
the tunnel splitting of these states is negligible in zero-field
($\sim$~Hz). However, it is possible to increase this splitting to
a practical range ($\sim$~GHz) with a transverse magnetic field.
While there remain technical challenges along the road map towards
molecule-based quantum devices (e.g. low operating temperatures,
methods for addressing nanometer-sized molecules, etc.), the
present study demonstrates that the "bottom-up" (molecular)
approach provides excellent opportunities to study coherent
quantum superposition states. Future materials design strategies
will, therefore, explore the following possibilities: optical
control of the exchange coupling between the two halves of a
dimer; increased isolation of the dimers in order to further
reduce decoherence; and the inclusion of some form of asymmetry
within the dimer (e.g. uncompensated electronic spins, or
selective nuclear spin labeling), thereby facilitating readout of
the state of the system.



\clearpage
\bibliographystyle{Science}


\begin{scilastnote}
\item We thank W. Wernsdorfer for useful discussion. This work was supported by the NSF and by Research Corporation.
\end{scilastnote}


\clearpage

\noindent {\bf Fig. 1.} The
[Mn$_4$O$_3$Cl$_4$(O$_2$CEt)$_3$(py)$_3$]$_2$ dimer; the Mn$^{3+}$
and Mn$^{4+}$ ions are labeled Mn and Mn$^\prime$, respectively.
The dashed lines represent the C$-$H$\cdot\cdot\cdot$Cl hydrogen
bonds holding the dimer together, and the dotted line is the close
approach of the central bridging Cl atoms believed to be the main
pathway for the exchange interaction between the two Mn$_4$
molecules.

\bigskip

\noindent {\bf Fig. 2.} Schematic showing the lowest energy states
(M$=-9$ to $-6$, not to scale) of the dimer: the zeroth order
energy levels and eigenvectors are shown on the left; energy
shifts due to the exchange bias are shown in the center; and the
results of a full quantum calculation (Eq.~\ref{Eq2}) are
displayed on the right. The colors denote the total angular
momentum (M) state of the dimer, and the levels have been numbered
to aid discussion. Several of the strongest EPR transitions are
indicated by arrows [(a) thru (g)]. The corrected eigenvectors are
listed next to each state: based on the deduced value of $J$ ({\em
vide infra}), $\beta=0.400$ and $\beta^\prime=0.231$; the
subscripts {\em S} and {\em A} (on a state $|m_1,m_2\rangle$)
respectively denote symmetric and antisymmetric combinations of
$|m_1,m_2\rangle$ and $|m_2,m_1\rangle$; and the $C_s$ are
normalization constants.

\bigskip

\noindent {\bf Fig. 3.} The left-hand panel displays temperature
dependent easy-axis data obtained for the [Mn$_4$]$_2$ dimer at
145~GHz (the dips in transmission correspond to EPR); the inset
(black trace) shows a single 6~K, 140~GHz, spectrum obtained for a
monomeric Mn$_4$ complex [the resonances are labeled according to
the $m$ states from which the transitions were
excited~\cite{HillMn4Poly}]. The right-hand panel contains
simulations of the dimer data, while the inset illustrates the
effect of the transverse part of the exchange ($J_{xy}$) for four
values of $J_{xy}/J_z$ (T$=8$~K). In both figures, resonances (a)
through (g) correspond to the labeled transitions in Fig.~2;
resonances ($x$), (h) and (i) are discussed in the main text. A
Gaussian distribution in $D$ ($\sigma_D\sim1\%$) was included in
the simulations in order to obtain realistic
lineshapes~\cite{Park1}.

\bigskip

\noindent {\bf Fig. 4.} A single fit to Eq.~\ref{Eq2} of the
positions of EPR peaks obtained at several frequencies. The
optimum Hamiltonian parameters were obtained from this fit. The
transitions have been labeled for comparisons with Figs.~2 and 3.

\clearpage
\begin{figure}
\includegraphics[width=0.9\textwidth]{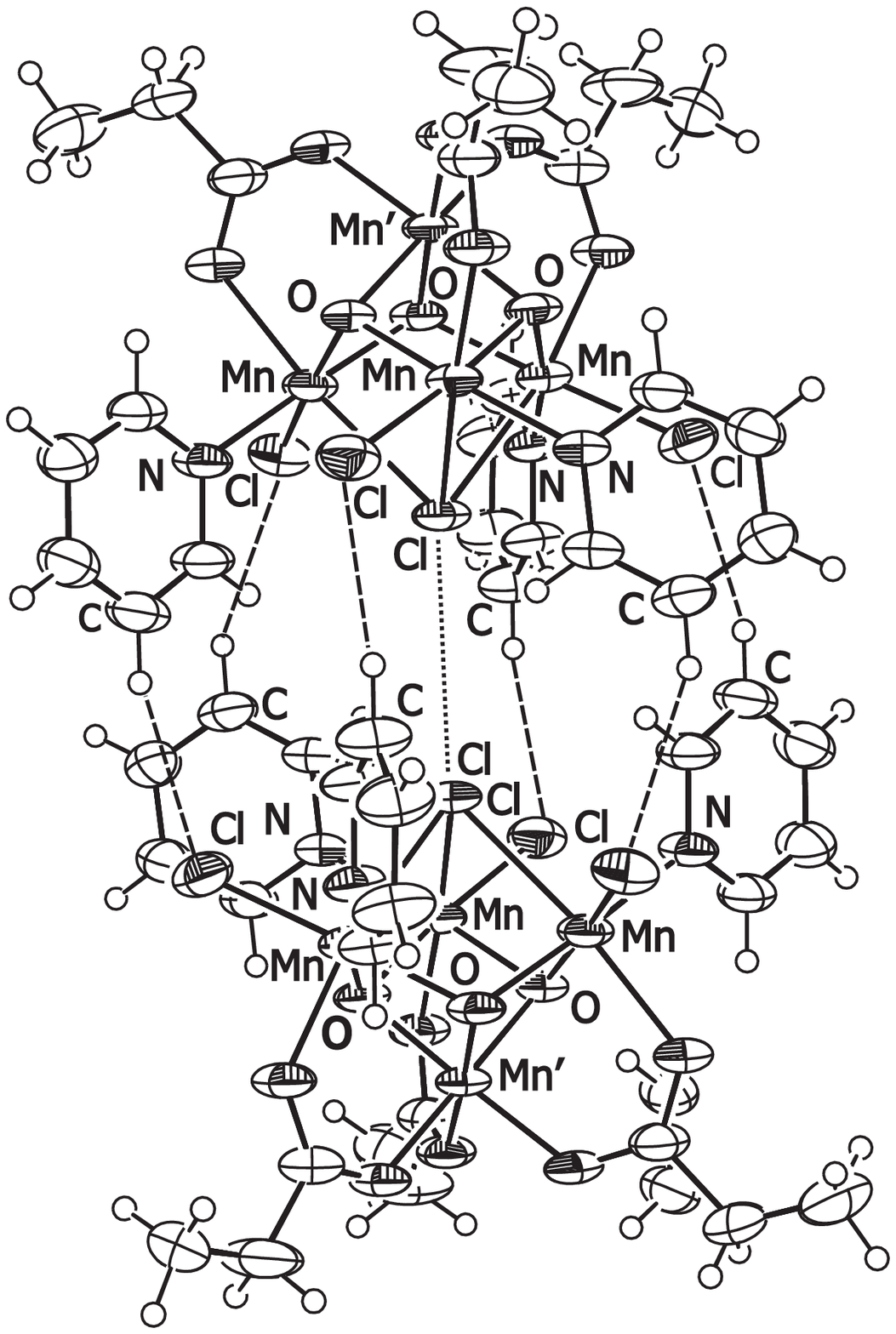}
\caption{\label{Fig1} S. Hill {\em et al.}}
\end{figure}
\clearpage

\clearpage
\begin{figure}
\includegraphics[width=1\textwidth]{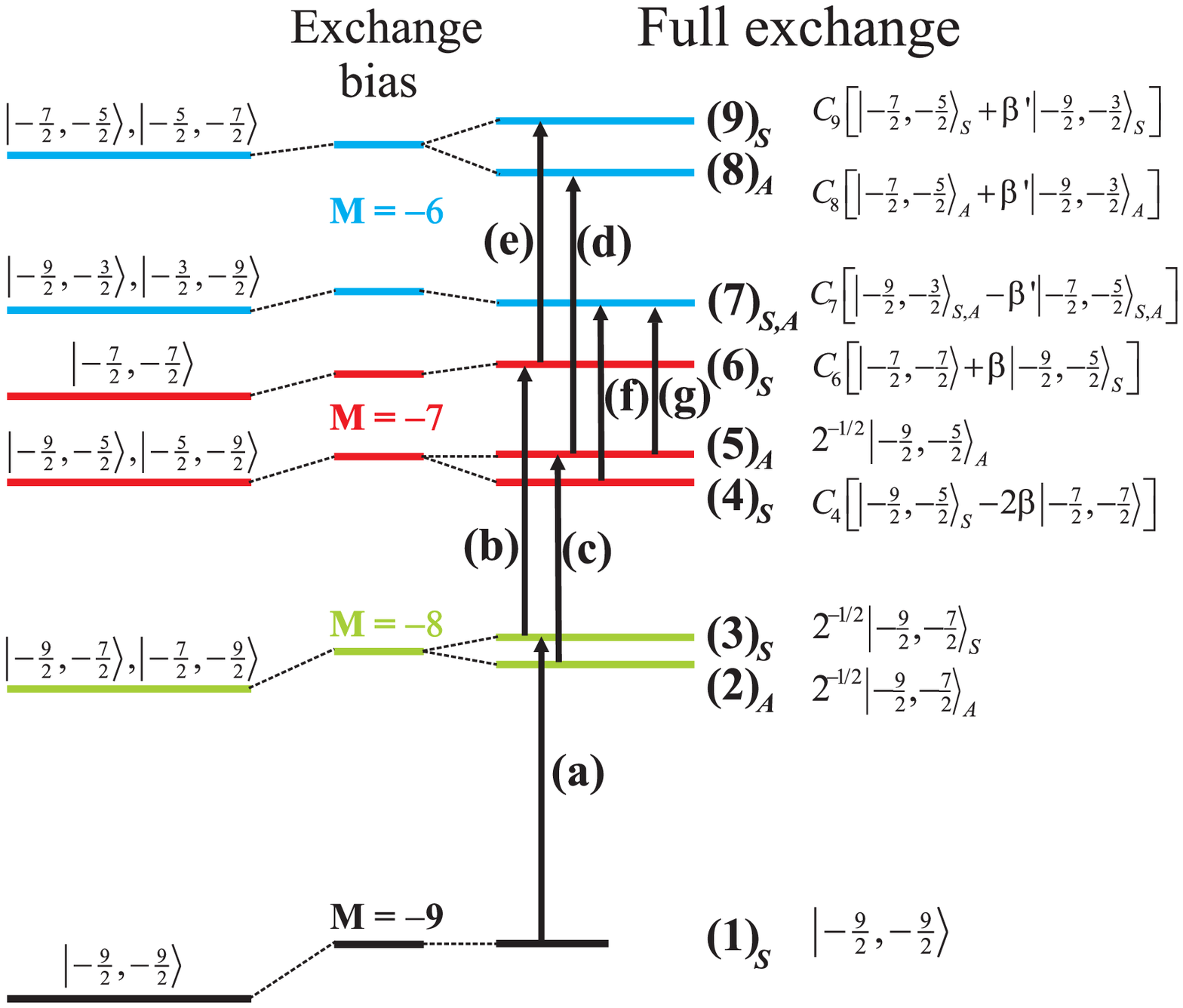}
\caption{\label{Fig2} S. Hill {\em et al.}}
\end{figure}
\clearpage


\begin{figure}
\includegraphics[width=1\textwidth]{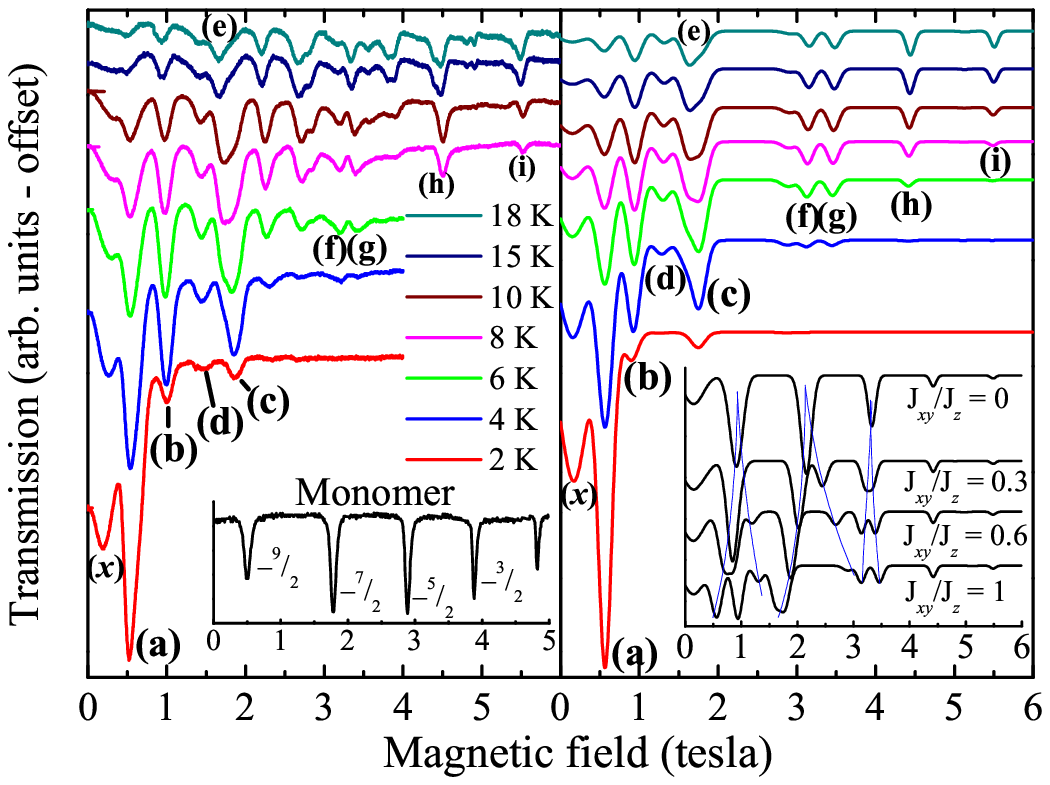}
\caption{\label{Fig3} S. Hill {\em et al.}}
\end{figure}
\clearpage

\begin{figure}
\includegraphics[width=1\textwidth]{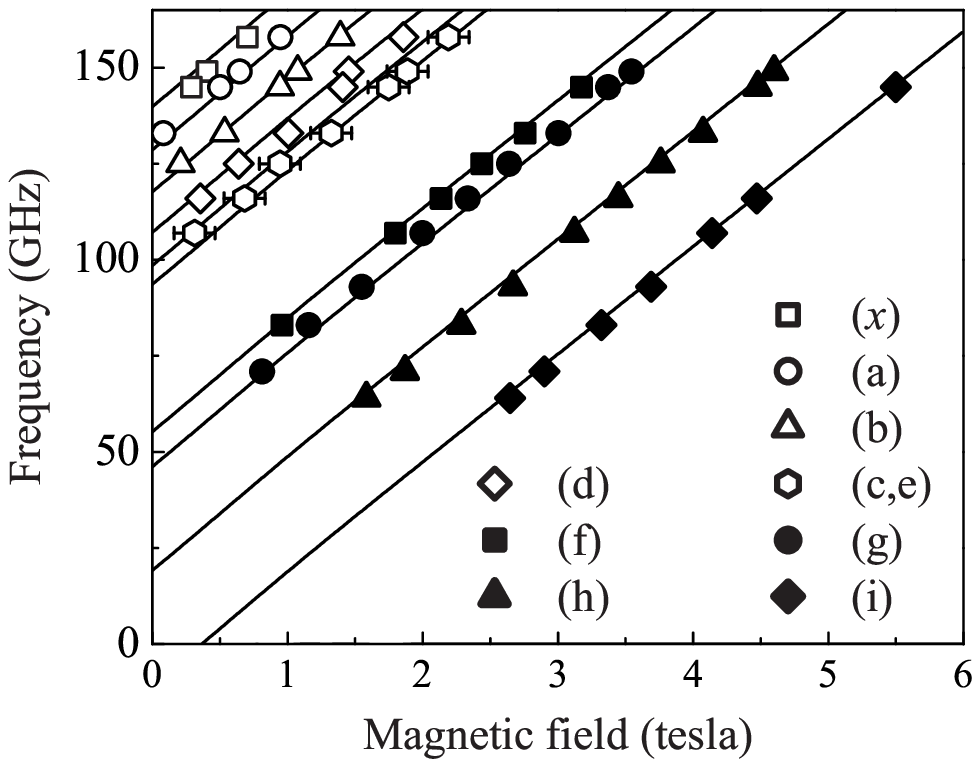}
\caption{\label{Fig4} S. Hill {\em et al.}}
\end{figure}

\end{document}